\documentclass[aps,prl,reprint]{revtex4-1}
\usepackage[ansinew]{inputenc}
\usepackage{graphicx}
\usepackage{textcomp}
\usepackage{amsmath}
\usepackage{hyperref}

\renewcommand{\figurename}{\textbf{Figure}}
\renewcommand{\thefigure}{\textbf{\arabic{figure}}}

\begin{document}

\title{A Single-Atom Quantum Memory}

\author{Holger P. Specht}
\email{Current address: Osram Opto Semiconductors GmbH, Leibnizstr. 4, 93055 Regensburg, Germany}
\author{Christian Nölleke}
\author{Andreas Reiserer}
\author{Manuel Uphoff}
\author{Eden Figueroa}
\author{Stephan Ritter}
\email[To whom correspondence should be addressed. Email: ]{stephan.ritter@mpq.mpg.de}
\author{Gerhard Rempe}
\affiliation{Max-Planck-Institut für Quantenoptik, Hans-Kopfermann-Strasse 1, 85748 Garching, Germany}

\maketitle

\textbf{The faithful storage of a quantum bit of light is essential for long-distance quantum communication, quantum networking and distributed quantum computing \cite{lvovsky2009}. The required optical quantum memory must, first, be able to receive and recreate the photonic qubit and, second, store an unknown quantum state of light better than any classical device. These two requirements have so far been met only by ensembles of material particles storing the information in collective excitations \cite{matsukevich2006,choi2008,tanji2009,jin2010,saglamyurek2011,clausen2011}. Recent developments, however, have paved the way for a new approach in which the information exchange happens between single quanta of light and matter \cite{cirac1997,wilk2007,boozer2007,togan2010}. This single-particle approach allows one to address the material qubit and thus has fundamental advantages for realistic implementations: First, to combat inevitable losses and finite efficiencies, it enables a heralding mechanism that signals the successful storage of a photon by means of state detection \cite{lloyd2001,bochmann2010,piro2011}. Second, it allows for individual qubit manipulations, opening up avenues for in situ processing of the stored quantum information. Here we demonstrate the most fundamental implementation of such a quantum memory by mapping arbitrary polarization states of light into and out of a single atom trapped inside an optical cavity. The memory performance is analyzed through full quantum process tomography. The average fidelity is measured to be 93\,\% and low decoherence rates result in storage times exceeding 180\,\textmu s. This makes our system a versatile quantum node with excellent perspectives for optical quantum gates \cite{jaksch1999,jaksch2000} and quantum repeaters \cite{briegel1998}.}

The robustness of a light qubit against inevitable photon loss during propagation and detection is essential for realistic applications. We therefore encode our qubit in the polarization degree of freedom. In this way, photon loss reduces a quantum protocol's efficiency, but cannot compromise its fidelity. Realizing such a polarization quantum memory has two main challenges, namely, finding a material qubit with a suitable energy-level scheme and achieving strong light-matter coupling for efficient and reversible information exchange. Both challenges are met in the system reported here. It features straightaway competitive fidelity and efficiency and longer storage times than all previous demonstrations \cite{matsukevich2006,choi2008,tanji2009,jin2010,saglamyurek2011,clausen2011}. Even better, clear strategies for improving its characteristics are identified. No fundamental physical limits have been reached, testifying to the usefulness of this approach and indicating a promising future for single-atom quantum memories.

\begin{figure}
\centering
\includegraphics[width=1\columnwidth]{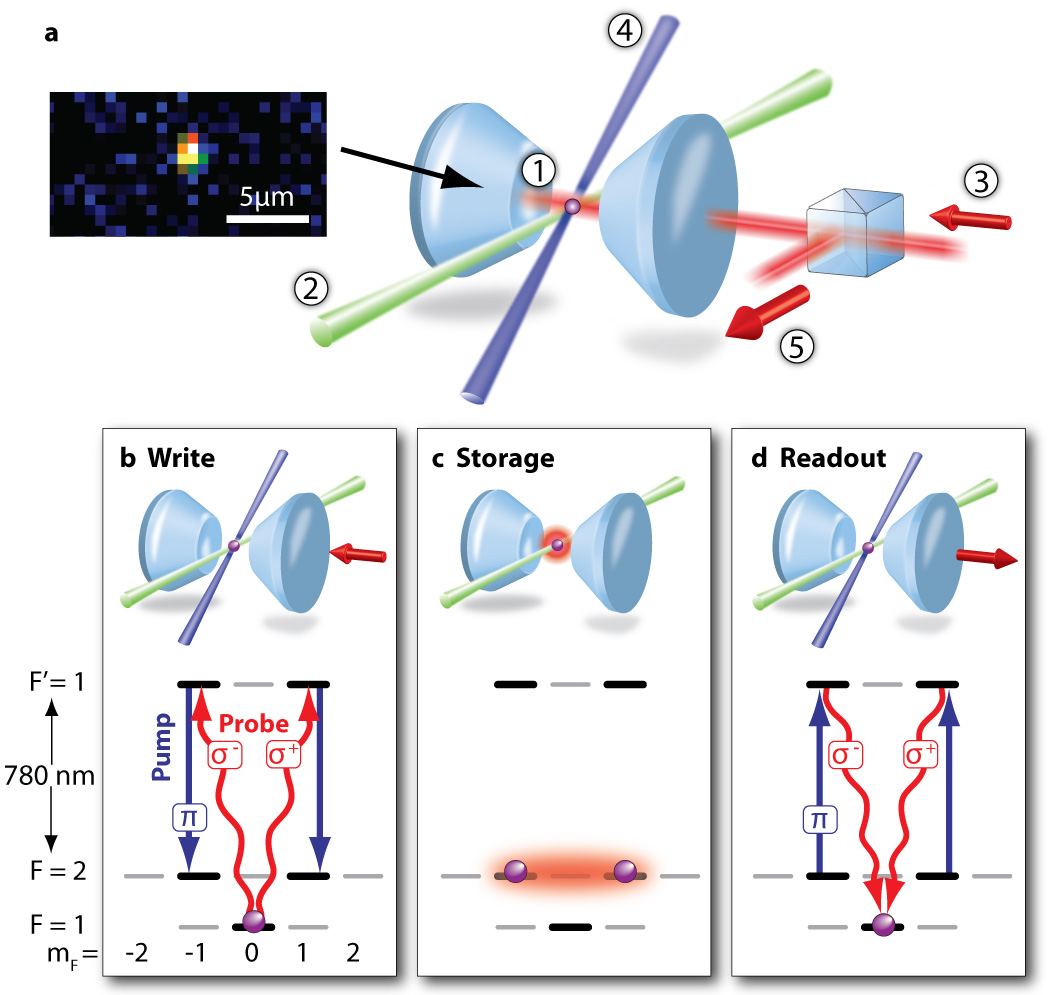}
\caption{\textbf{Single-atom quantum memory.} \textbf{a}, A single atom (1) (the inset shows a typical fluorescence image) is trapped at the centre of a high-finesse optical cavity using a far-detuned standing-wave dipole trap (2). \textbf{b}, An impinging weak coherent pulse (3) with arbitrary polarization is converted into an atomic spin excitation using a $\pi$-polarized pump laser (4). \textbf{c}, This maps the photonic polarization qubit onto a long-lived superposition of the $|F\!=\!2,m_F\!=\!\pm1\rangle$ ground states of the atom. \textbf{d}, After a variable storage time, the polarization qubit is retrieved by the production of a single photon (5).}
\end{figure}

The hardware of our quantum memory is a single $^{87}$Rb atom quasi-permanently trapped at the centre of a high-finesse optical cavity (Fig.\,1) (see Supplementary Information)\cite{hijlkema2007}. The efficiencies for coupling photons into and out of the cavity are optimized by using one high-reflectivity mirror (transmission $T<6$\,ppm) and an output coupler with higher transmission ($T \approx 100$\,ppm). The atom-cavity coupling constant $g$ is comparable to the cavity-field decay rate $\kappa$ and the atomic polarization decay rate $\gamma$ of $^{87}$Rb, and therefore the system is in the intermediate coupling regime of cavity QED: $(g, \kappa, \gamma) = 2\pi \times(5, 2.5, 3)$\,MHz.

The memory is initialized by preparing the atom in the state $|F\!=\!1, m_F\!=\!0\rangle$ with an efficiency higher than 90\,\%. It is probed using weak coherent pulses (mean photon number $\bar{n} < 1$) resonant with the cavity and 12\,MHz blue detuned from the Stark shifted $F\!=\!1 \leftrightarrow F'\!=\!1$ resonance of the D$_2$ line. Simultaneously with the arrival of the incoming pulse, the Rabi frequency of a control laser in Raman-resonance with the cavity is adiabatically ramped down to zero. The control laser is oriented perpendicular and is $\pi$-polarized with respect to the quantization axis defined by the cavity axis. This transfers the atomic population to the ground states $|F\!=\!2,m_F\!=\!\pm 1\rangle$ through a cavity-mediated stimulated Raman adiabatic passage (STIRAP) \cite{hennrich2000,cirac1997} (Fig.\,1a). During this coherent process, the phase relation between the $\sigma^{\pm}$  input polarization modes is mapped to a relative phase between the populations of the aforementioned Zeeman substates (Fig.\,1b). Efficient storage is achieved by optimizing the centre and width of the falling edge of the control laser pulse following a $\cos^2$ function in time.

After a variable storage time, a second control laser pulse converts the atomic qubit back onto the polarization of a produced single photon \cite{wilk2007}, which ideally will be in the same polarization state as the initial incoming pulse (Fig.\,1c). The wave-packet shape of the retrieved photon can thereby be adjusted, independent of the form of the input pulse, which is of great importance in more complex quantum networks. A non-polarizing beam splitter is used to separate the paths for incoming and outgoing photons, with the latter being directed to a detection setup, where the polarization state of the photon can be analyzed in any basis.

\begin{figure}
\centering
\includegraphics[width=\columnwidth]{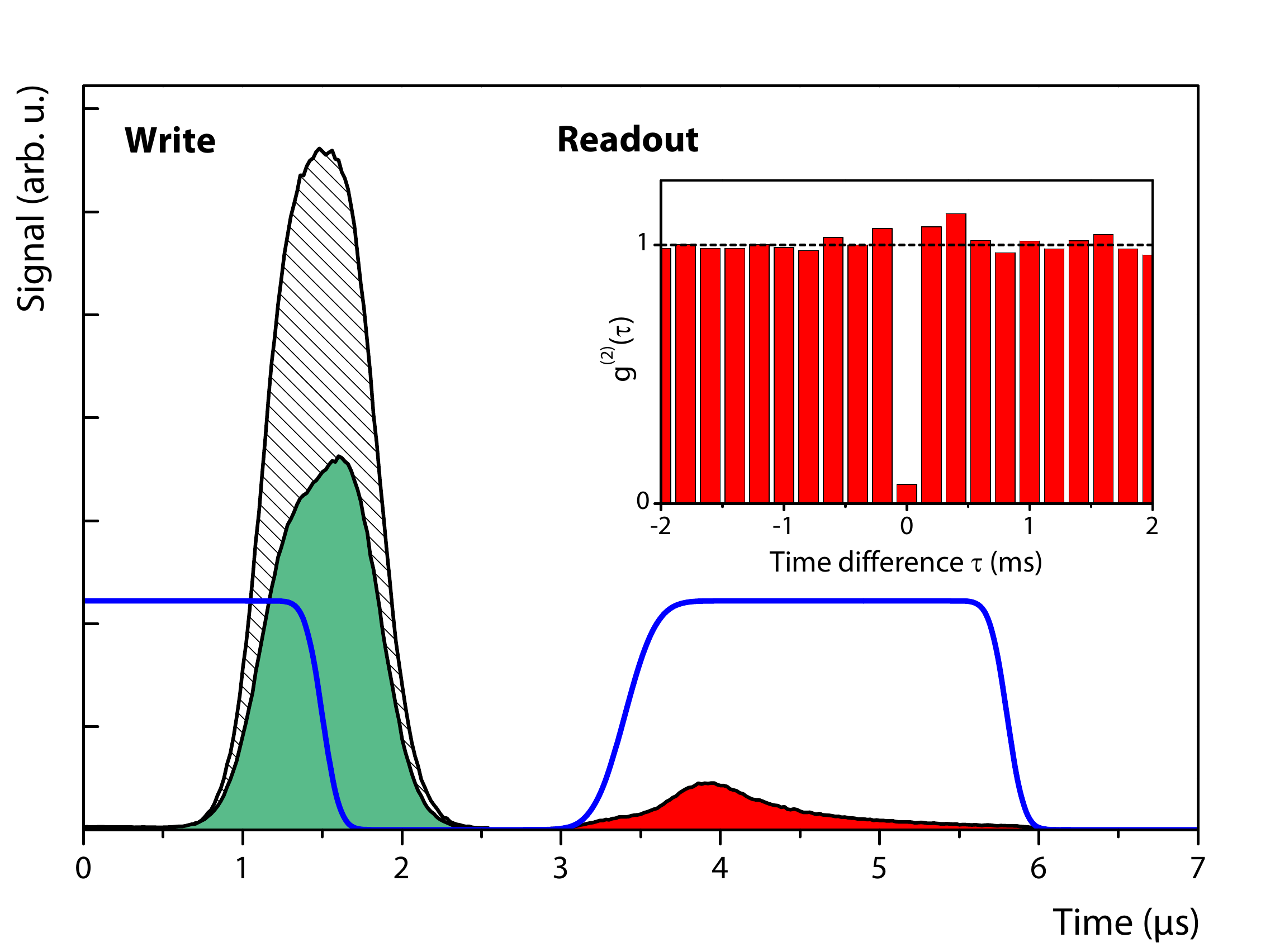}
\caption{\textbf{Write and read process of the memory.}
While the input photon pulse (shaded area, full width at half maximum 0.7\,\textmu s) impinges on the cavity, the power of the control laser (blue, curve only schematic) is adiabatically ramped down to zero. Part of the incident light is directly reflected from the memory (green area). After an adjustable storage time, a photon is produced by adiabatically switching on the control laser. The area of the retrieved photon pulse (red) relative to the incident pulse yields the overall storage efficiency of the memory of 9.3\,\%. The inset shows the correlation function $g^{(2)}(\tau)$ of the retrieved photons as measured by a Hanbury Brown--Twiss experiment. The ratio of two-photon to single-photon events is 0.5\,\% and can be fully explained by stray light and dark counts of the detectors. This verifies the single-photon character of the retrieved light.}
\end{figure}

We define the efficiency of our quantum memory as the fraction of photons retrieved from the cavity after storage, normalized to the input (Fig.\,2). It was measured to be $(9.3\pm 1)\,\%$ (see Supplementary Information). The uncertainty is dominated by fluctuations in the system parameters (e.g.\ the atomic position). We independently measured the photon production efficiency which is relevant for the readout process to be 56\,\%. The efficiency is clearly limited by the atom-cavity coupling \cite{muecke2010,gorshkov2007b} and could be increased considerably by better localization of the atom or a smaller cavity mode volume. Due to the single-atom character of our quantum memory, only single photons can be stored, even if the input is a coherent state. This is confirmed by the measured antibunching of the retrieved photons (inset in Fig.\,2).

We have characterized our quantum memory with six different input polarizations: $|R\rangle$, $|L\rangle$, $|H, V\rangle=(|R\rangle \pm |L\rangle)/\sqrt{2}$ and $|D, A\rangle = (|R\rangle \pm i|L\rangle)/\sqrt{2}$. For each input state, the output is analyzed in three orthogonal bases, which allows for a full reconstruction of the retrieved photon's density matrix \cite{james2001}. The fidelity is defined as the overlap of the density matrix $\rho$ of the measured output with the ideal input state $|\psi_i\rangle$: $F=\langle\psi_i | \rho | \psi_i \rangle$. It therefore not only includes the storage process but also possible infidelities of the read-out. The fidelities of the six input states for a storage time of 2\,\textmu s are
\begin{align*}
&&|H\rangle:\ 92.2\,(4)\,\%,\quad &|V\rangle:\ 92.0 (4)\,\%\,,& \\
&&|D\rangle:\ 91.9\,(5)\,\%,\quad &|A\rangle:\ 90.9 (4)\,\%\,,& \\
&&|R\rangle:\ 95.1\,(4)\,\%,\quad &|L\rangle:\ 94.2 (4)\,\%\,,&
\end{align*}
yielding an arithmetic mean \cite{bowdrey2002} of $92.7 (2)$\,\%. The average fidelity of the input states directly reflected from the cavity is $99.20 (2)$\,\%, proving that errors during preparation, propagation and detection of the photon's polarization state are small.

Our fidelity has to be compared to what can be achieved following classical strategies. Unlike a quantum memory, a classical memory effectively performs a measurement on the state of the incoming qubit. Based on the incomplete information gained, it will try to reproduce the input state, a procedure known as an {\it intercept-resend attack} in the context of quantum key distribution \cite{curty2005}. The maximum average fidelity that can be achieved with such a classical memory and single input photons is $2/3$ \cite{massar1995}. When tested with coherent light, a classical memory could gain additional information by measuring more than one photon. In that case, the fidelity threshold for a quantum memory with our characteristics is increased to 80\,\% (see Supplementary Information). Our fidelity is well above both of these thresholds, thereby proving the quantum nature of the memory.

In order to obtain the storage fidelity for an arbitrary input state, the process can be further analyzed using quantum process tomography \cite{chuang1997,nielsen2000,poyatos1997}. Every polarization state can be represented by a three-dimensional vector $\vec{S}$. Pure states have $|\vec{S}|=1$ and therefore lie on the unit Poincaré sphere. An intuitive representation for the infidelity of the storage process for any input state is given by the deformation of this sphere (Fig.\,3a).

\begin{figure}
\centering
\includegraphics[width=\columnwidth]{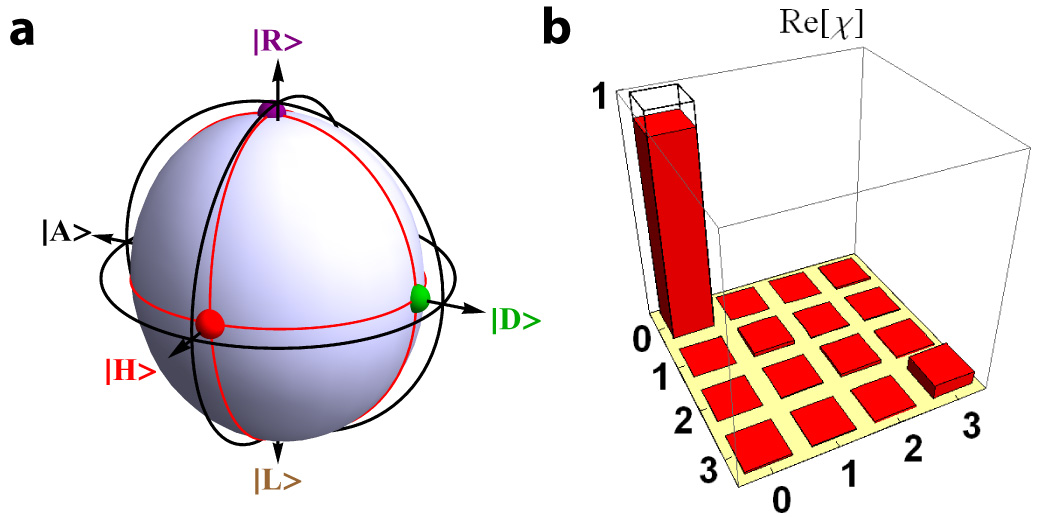}
\caption{\textbf{Tomography of the storage process for a storage time of 2\,\textmu s.} \textbf{a}, The storage process can be depicted as a deformation (bluish ellipsoid) of the unit Poincaré sphere (black circles). The coloured dots indicate the polarizations of the input states on the axes (same colour coding) after storage. The average fidelity is $F=92.7\pm0.2$\,\%, clearly proving the quantum character of the memory. \textbf{b}, Real part of the process matrix $\chi$ as obtained from a full quantum process tomography. All imaginary parts are close to zero (largest magnitude 0.034).}
\end{figure}

An alternative representation for the quantum process is obtained via the process matrix $\chi$. It maps an input density matrix $\rho_\mathrm{in}$ onto the corresponding output state $\rho_\mathrm{out}$, via
$$\rho_\mathrm{out} = \sum_{m,n=0}^3 \chi_{mn} \sigma_m \rho_\mathrm{in} \sigma^\dagger_n .$$
Here, the operators $\sigma_i$ form a set of Pauli matrices. As we observe no systematic dependence of the efficiency on the polarization of the input pulse, the efficiency can be excluded by normalizing the density matrices. Then, for a process preserving the input state, $\chi_{00}$ is equal to one, while all other elements are zero. As shown in Fig.\,3b, the main deviation of our $\chi$ matrix from the ideal one is a nonzero $\chi_{33}$, indicating dephasing between the Zeeman states.

Several processes contribute to the infidelity of our quantum memory. The main contribution arises from the production of a photon independent of the state of the input photon, caused by non-optimal optical pumping to $F\!=\!1$. We have confirmed that this occurs with an efficiency of less than 1.3\,\% by measuring the efficiency of producing a photon without incoming photon during the storage process. The detrimental effect of incoherent transfer of the atomic population \cite{boozer2007} is minimized by working in two-photon resonance but 12\,MHz off-resonant from the atomic transition. Stray light and dark counts only occur with a probability of 0.3\,\% relative to the number of incoming photons.

We have also characterized our memory by measuring the write-read fidelity as a function of the storage time (Fig.\,4a). The classical limit of $2/3$ is reached after 82\,\textmu s. For longer storage times, the system approaches $F=50\,\%$ as expected for a completely mixed output state. Two main aspects contribute to the observed decay: residual fluctuating magnetic fields and differential Stark shifts caused by the trapping lasers. Both lead to a dephasing of the stored state. Gaussian field fluctuations explain the observed Gaussian decay of fidelity with storage time. We find the fidelity for circularly polarized light to decay slower than for linear polarizations, as the respective atomic states are eigenstates of any longitudinal magnetic field. By applying a guiding magnetic field $B_\mathrm{guide}$, the storage times can be further increased (Fig.\,4b). Using a field parallel to the cavity axis suppresses the influence of perpendicular fluctuations $\Delta B$ on the absolute field stability by a factor $\Delta B/2B_\mathrm{guide}$. The influence of longitudinal fluctuations, however, is unaffected. The guiding field causes the relative phase between the states $|F\!=\!2,m_F\!=\! \pm 1\rangle$ to evolve at twice the Larmor frequency. Since linear polarizations are stored as superpositions of these states, the measured fidelity shows a sinusoidal oscillation. Knowing the precession period of 21\,\textmu s for a guiding field of $B_\mathrm{guide} = 34$\,mG and the storage time, this can be compensated using polarization optics. The slightly reduced fidelity at short storage times is caused by an already significant Larmor precession during the storage and retrieval process. The overall results, however, are increased fidelities for a given storage time and therefore longer overall storage times. This is evidenced by a more than two-fold increase in storage time, with the average fidelity reaching the classical limit of $2/3$ after 184\,\textmu s. The effect for purely circular polarizations, which are stored in eigenstates of the guiding field, is even more dramatic, with the fidelity staying almost constant during the observation period.

\begin{figure}
\centering
\includegraphics[width=\columnwidth]{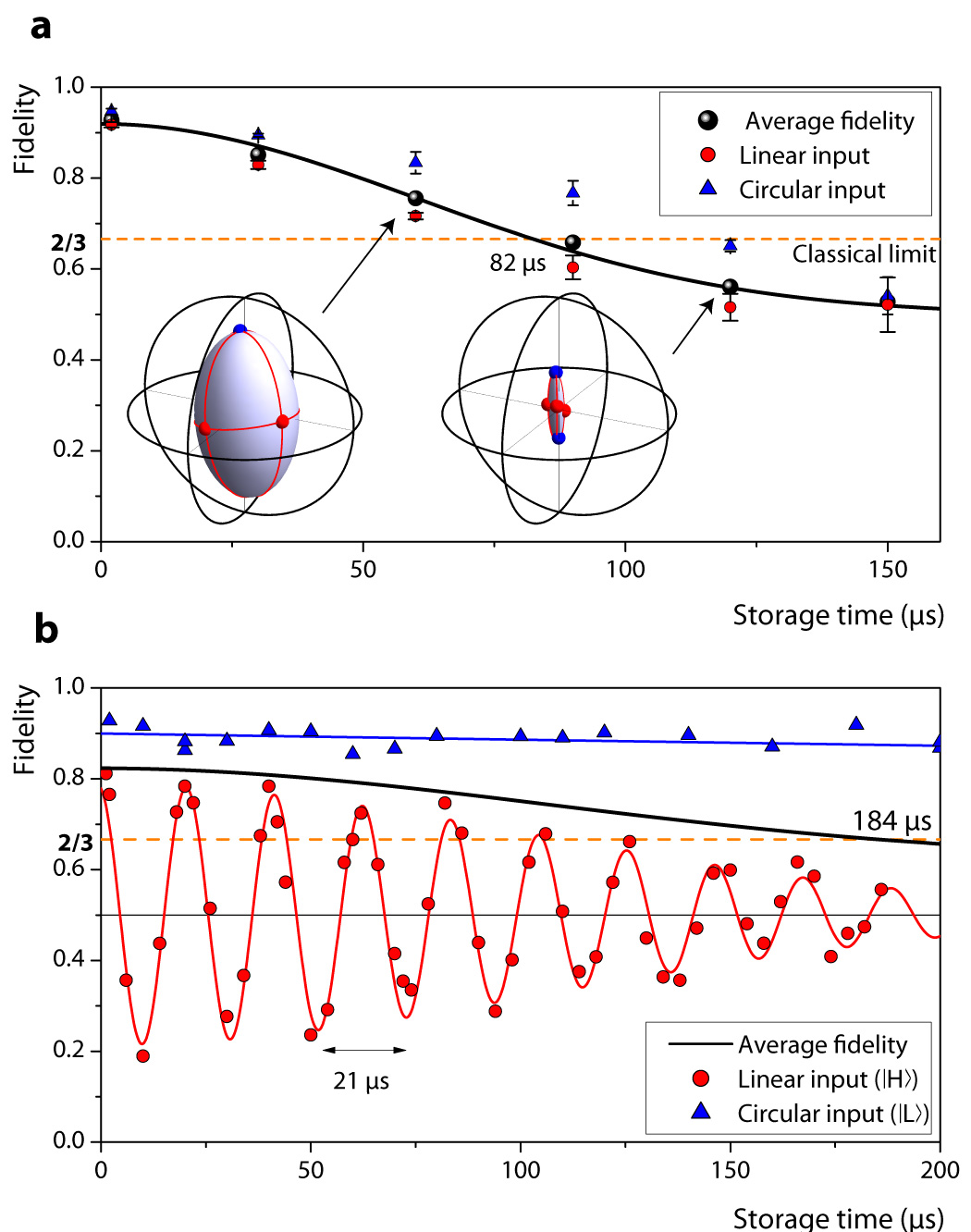}
\caption{\textbf{Storage time.}
\textbf{a}, The fidelity between ideal input and measured output state is plotted as a function of the storage time. A significantly slower decrease for the circular polarizations (blue triangles) compared to all linear polarizations (red circles) can be observed. The error bars denote the scatter for individual polarizations. For the average fidelity (black dots and black Gaussian fit), the classical limit of $2/3$ is reached after 82\,\textmu s. For the retrieved photon, the evolution of the Poincaré sphere in Stokes space confirms that the main decoherence mechanism is dephasing between the Zeeman states that encode $|R\rangle$ and $|L\rangle$.
\textbf{b}, Increased storage times can be obtained by applying a constant guiding field of $B_\mathrm{guide} = 34$\,mG along the cavity axis. The resulting Larmor precession of the atomic Zeeman coherence results in an oscillation of the retrieved polarization for linearly polarized states (only $|H\rangle$ is shown) with a period of 21\,\textmu s. The net effect is a stabilization of the coherence of the atomic qubit and an increase in the average achievable storage time by more than a factor of 2. In \textbf{b} the average fidelity is defined as the mean of the fidelities for $|H\rangle$ (doubly weighted) and $|L\rangle$.}
\end{figure}

In summary, we have demonstrated a quantum memory for polarization qubits based on a single trapped atom with an average fidelity of 93\,\%, storage times of 184\,\textmu s and an overall efficiency of 9.3\,\%. Further improvements are possible by storing the qubit in a decoherence-free subspace, with the prospect of storage times exceeding several seconds \cite{langer2005}. The single-particle nature of our memory also provides the potential for the implementation of a herald. As a successful storage attempt depopulates the atomic $F\!=\!1$ state, its use as the bright state in hyperfine state detection \cite{bochmann2010} will herald the storage without degradation of the qubit \cite{lloyd2001}. Such a heralded memory can be used to perform a quantum non-demolition measurement on the presence of a photonic qubit.
Furthermore, the demonstrated mapping of a photon state onto an atom can be used to realize a deterministic entanglement scheme between remote atomic qubits: If a first emitting atom is entangled with a photon \cite{wilk2007}, the absorption of this photon in a memory atom directly produces entanglement between the two atoms \cite{cirac1997}. This establishes single-atom cavity systems as universal nodes of a future quantum network.

\begin{acknowledgments}
We thank N. Kiesel for helpful discussions and A. Neuzner for experimental assistance. This work was supported by the Deutsche Forschungsgemeinschaft (Research Unit 635), by the European Union (Collaborative Project AQUTE) and by the Bundesministerium für Bildung und Forschung via IKT 2020 (QK\_QuOReP). E. F. acknowledges support from the Alexander von Humboldt Foundation.
\end{acknowledgments}

\onecolumngrid
\vspace{70pt}
\section{Supplementary Information}
\quad
\twocolumngrid

\section*{System preparation and experimental sequence}
The scheme starts with the probabilistic loading of a single atom from a magneto-optical trap. A loading event takes 1--2 seconds. The system switches to trapping mode (2\,W standing-wave dipole laser beam at 1064\,nm, focused to a waist of 16\,\textmu m at the centre of the cavity) when a single atom is detected. The innermost part of the cavity region is imaged onto an EM-CCD camera using an objective with high-numerical aperture ($\mathrm{NA}=0.43$). Every second the position of the atom is determined and, in case of a displacement, the standing-wave pattern of the dipole trap is shifted by means of a rotatable glass-plate in front of the retroreflecting mirror. This allows for micro-positioning of the atom along the dipole trap axis exactly at the centre of the cavity mode [30]. The standing-wave dipole trap (3\,mK trapping potential, 133\,MHz maximum AC-Stark shift) causes a spatially modulated atomic transition frequency. This allows for Sisyphus-like cooling via a combination of a lin $\perp$ lin retroreflected laser resonant with the unshifted $F\!=\!2 \leftrightarrow F'\!=\!3$ cycling transition and a repumping laser on the $F\!=\!1 \leftrightarrow F'\!=\!2$ transition [31]. The resulting average trapping time during the storage sequence is 70\,s. As this is much longer than the loading period, single atoms are quasi-permanently available.

In order to characterize the performance of the quantum memory, the following experimental sequence is used: Initially, the atom is cooled for 150\,\textmu s employing the above-mentioned cooling mechanism. Subsequently, the reflection of a coherent input pulse off the cavity is measured as a reference, while the atom in the $F\!=\!2$ state is effectively decoupled from the cavity. Then, the atom is transferred to the state $|F\!=\!1, m_F\!=\!0\rangle$ using a two-step process. First, it is pumped to the $F\!=\!1$ manifold using a stimulated Raman adiabatic passage (STIRAP). In a second step, a $\pi$-polarized laser resonant with the $F\!=\!1\leftrightarrow F'\!=\!1$ transition transfers the population to $|F\!=\!1,m_F\!=\!0\rangle$. This procedure is repeated five times to achieve a pumping efficiency above 90\,\%. Subsequently, an impinging coherent pulse is written into the memory, stored for a variable time and read out. The whole sequence is repeated at a rate between 2.5 and 5\,kHz, depending on the storage time.

\section*{Efficiencies}
\subsection*{Definition of the storage efficiency}
For the case of perfect single photon input pulses an intuitive definition for the write-read efficiency of the quantum memory is the fraction of events in which a photon is retrieved. Using an optimized impedance matching, nearly unity absorption efficiency is in principle possible [32]. To extend this definition to the cases of probabilistic single photons and coherent input pulses, events with vacuum states and higher photon numbers have to be accounted for.

A possible definition for the write-read efficiency is the probability of success per non-empty input pulse, i.e. the ratio of successful storage and retrieval events relative to all events with at least one photon in the input pulse. However, this definition only yields an upper bound for the single-photon efficiency: If several photons are available for storage, as is possible for coherent pulses, the efficiency will be higher than for single photons.

A definition of the write-read efficiency more suitable for coherent pulses is the ratio of the mean energy of the retrieved and impinging pulse. It is used throughout here. As the memory can store only up to one photon, this definition gives a lower bound for the single-photon efficiency. For example, a perfect quantum memory tested with coherent input pulses with on average one photon ($\bar{n} = 1$) will yield an efficiency of 63\,\% under this definition.

\subsection*{Photon detection efficiency}
Besides the non-unity efficiency of the quantum memory, photon loss decreases the detected signal of the impinging and retrieved pulse. Storage and retrieval of the photon is performed through the outcoupling mirror. An intracavity photon leaves the cavity through this mirror with a probability of 92\,\%. The cavity mode on this side is spatially mode-matched to a single-mode optical fiber. The overall detection efficiency for a photon at the output of the cavity is 41\,\%. This is a combination of the fiber coupling efficiency (86\,\%), transmission through optical elements (96\,\%) and the quantum efficiency of the single photon counting modules (50\,\%). These inefficiencies are identical for all photons coming from the cavity, i.e. for both, reflected as well as retrieved photons. They therefore cancel out in the calculation of the efficiency.

In the experiment, the input light reflected from the empty, resonant cavity is monitored. To deduce the energy of the input pulse, this value has to be corrected for cavity-associated losses, i.e. absorbtion and scattering of light inside the cavity and partial transmission. Only 71\,\% of the light that is reflected off the non-resonant cavity is reflected off the resonant cavity. This was accounted for in Fig.\,2 of the main text. To retrieve the curve for the impinging light (shaded area), the measured reflection of the input pulse from the locked cavity was divided by 0.71. In addition, there is a difference in the detection efficiency for incoming light reflected off the unlocked cavity compared to light that leaves the cavity (e.g.\ light in the retrieved pulse). The former was determined to be 87\,\% of the latter. This is included in Fig.\,2 of the main text by dividing the reflected signals (shaded and green area) by 0.87. In total, the efficiency given in the main text is the fraction of retrieved photons relative to the number of photons impinging on the cavity, including imperfect mode matching of the input beam.

\renewcommand{\figurename}{\textbf{Supplementary Figure}}
\renewcommand{\thefigure}{\textbf{\arabic{figure}}}
\setcounter{figure}{0}
\begin{figure*}
\centering
\includegraphics[width=1.0\textwidth]{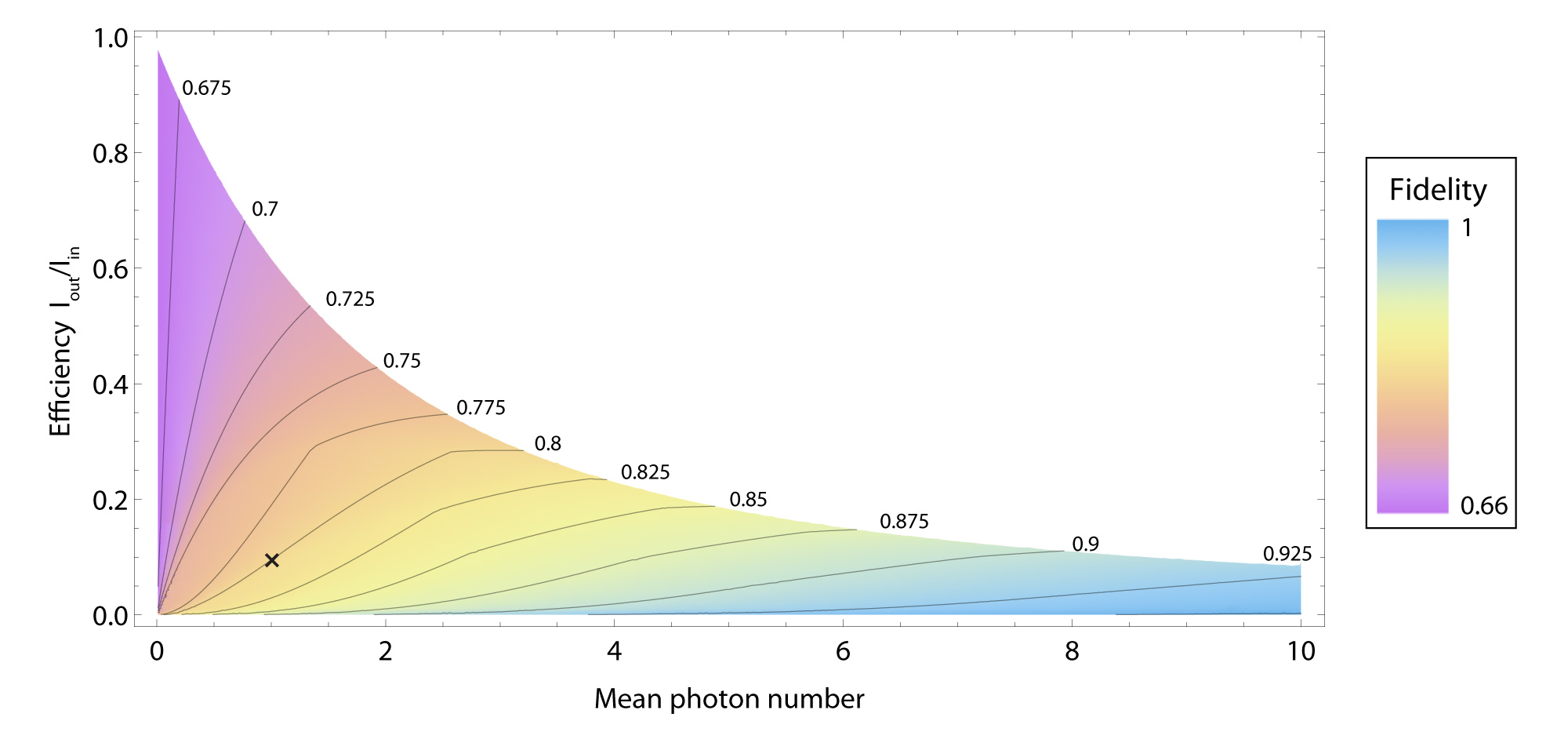}
\caption{\textbf{Fidelity threshold for classical memories.} The maximum average fidelity that is achievable with classical measurements is shown. It depends on the mean photon number of the coherent input pulse and on the required efficiency of the memory. The parameters relevant for our work are marked with a cross.}
\end{figure*}


\section*{Classical memories and coherent input pulses}
A classical memory performs a measurement on a qubit, stores the result classically and accomplishes readout by preparing a new qubit based on the obtained measurement result. In the context of quantum key distribution, this is known as an {\it intercept-resend attack} [33, 34]. The maximum average fidelity that can be achieved with such a classical memory and single input photons is $2/3$ (see [35]). If the input pulse consists of several identical photons, a classical memory could gain additional information by individual measurements of the photons in different bases [36]. For input pulses with fixed photon number the maximum achievable fidelity is [35]
\begin{equation}
\label{equ:fakeMemory}
\mathcal{F}_\mathrm{MP}(N) = \frac{N+1}{N+2}\,.
\end{equation}
In the case of coherent input pulses with a mean photon number $\bar{n}$, the number of photons per pulse $N$ follows a Poissonian distribution
\begin{equation}
\label{equ:Poisson}
p(\bar{n},N)=\frac{\bar{n}^N}{N!}e^{-\bar{n}}\,.
\end{equation}
The maximally achievable average fidelity can in this case be calculated as
\begin{equation}
\label{equ:fakeMemory_coherent}
\mathcal{F}_{coh}(\bar{n}) = \sum_{N\geq 1} \mathcal{F}_\mathrm{MP}(N)\frac{p(\bar{n},N)}{1-p(\bar{n},0)}\,.
\end{equation}
Thus, the fidelity threshold that discriminates between a classically possible process and a true quantum memory has to be increased. As long as this is properly accounted for, a quantum memory can unambiguously be characterized with coherent input pulses. As an example, a classical memory that is tested with coherent states with $\bar{n}=1$ can achieve a fidelity of 70.9\,\%.

An even more elaborate classical memory could also simulate non-unity efficiencies to increase the fidelity. The strategy of such a memory would be to produce no output when the input photon number is low, therefore only emitting a photon in cases of high enough incident photon numbers. For the user this would appear as an increased fidelity at the expense of a decreased efficiency of the memory. In the limit of extremely low efficiencies, fidelities approaching 100\,\% are possible.

Supplementary Fig.\,1 shows how the maximally achievable fidelity of a classical memory depends on its efficiency and the mean photon number of the coherent probe pulses exploiting both aforementioned strategies. Owing to our definition of the efficiency and no more than a single photon in the output, there is an upper limit for the efficiency at a given mean photon number (signaled by the boundary to the white region in Supplementary Fig.\,1). It starts at one for extremely low mean photon numbers and drops for large values as $1/N$. For our memory efficiency of 9.3\,\% and on average one input photon the fidelity threshold for proving a quantum memory is 80.1\,\% (black cross in Supplementary Fig.\,1).

\section*{Animations}
Supplementary Movie 1a and 1b (distributed separately) contain the temporal evolution of the storage process as a movie. Polarization measurements on the retrieved signal were interpolated for a sequence of storage times. While the magnetic field was minimized for the measurements of Supplementary Movie 1a (one measurement every 30\,\textmu s), a guiding field of 34\,mG was applied for Supplementary Movie 1b, leading to a rotation of the Poincaré sphere at twice the Larmor frequency (one measurement every 2--4\,\textmu s).

\section*{References for Supplementary Information}
\noindent
{[30]} Nußmann,~S. {\it et~al.\/} Submicron positioning of single atoms in a microcavity. {\it Phys. Rev. Lett.\/} {\bf 95}, 173602 (2005).\\
{[31]} Weber,~B. {\it et~al.\/} Photon-photon entanglement with a single trapped atom. {\it Phys. Rev. Lett.\/} {\bf 102}, 030501 (2009).\\
{[32]} Fleischhauer,~M., Yelin,~S.~F. \& Lukin,~M.~D. How to trap photons? Storing single-photon quantum states in collective atomic excitations. {\it Opt. Commun.\/} {\bf 179}, 395--410 (2000).\\
{[33]} Curty,~M. \& Lütkenhaus,~N. Intercept-resend attacks in the Bennett-Brassard 1984 quantum-key-distribution protocol with weak coherent pulses. {\it Phys. Rev. A\/} {\bf 71}, 062301 (2005).\\
{[34]} Félix,~S., Gisin,~N., Stefanova,~A. \& Zbinden,~H. Faint laser quantum key distribution: Eavesdropping exploiting multiphoton pulses. {\it Journal of Modern Optics\/} {\bf 48}, 2009--2021 (2001).\\
{[35]} Massar,~S. \& Popescu,~S. Optimal extraction of information from finite quantum ensembles. {\it Phys. Rev. Lett.\/} {\bf 74}, 1259--1263 (1995).\\
{[36]} Dus\v{e}k,~M., Jahma,~M. \& Lütkenhaus,~N. Unambiguous state discrimination in quantum cryptography with weak coherent states. {\it Phys. Rev. A\/} {\bf 62}, 022306 (2000).

\end{document}